\documentclass[a4paper,fleqn,usenatbib]{mnras}

\usepackage{newtxtext,newtxmath}

\usepackage[T1]{fontenc}
\usepackage{ae,aecompl}
\usepackage{pdfpages}

\usepackage{graphicx}	
\usepackage{amsmath}	
\usepackage{amssymb}	

\usepackage{bm}

\newcommand{\V}{\bm{v}}

\hypersetup{draft}   

\title[Avoiding Resonances]{Multiplanet systems in inviscid discs can avoid forming resonant chains}

\author[C.~P.~McNally et al.]{
Colin P.~McNally,$^{1}$\thanks{E-mail: c.mcnally@qmul.ac.uk (CPM)}
Richard P.~Nelson$^{1}$
and Sijme-Jan Paardekooper $^{1,2}$
\\
$^{1}$Astronomy Unit, School of Physics and Astronomy, Queen Mary University of London, London E1 4NS, UK\\
$^{2}$DAMTP, University of Cambridge, Wilberforce Road, Cambridge CB3 0WA, UK\\
}

\date{Accepted XXX. Received YYY; in original form ZZZ}

\pubyear{XXX}

\begin{document}
\label{firstpage}
\pagerange{\pageref{firstpage}--\pageref{lastpage}}
\maketitle

\begin{abstract}
Convergent migration involving multiple planets embedded in a viscous protoplanetary disc is expected to produce a
chain of planets in mean motion resonances, but the multiplanet systems observed by the Kepler spacecraft are generally 
not in resonance.
We demonstrate that under equivalent conditions, where in a viscous disc convergent migration
will form a long-term stable system of planets in a chain of mean motion resonances, 
migration in an inviscid disc often produces a system which is highly dynamically unstable.
In particular, if planets are massive enough to significantly perturb the disc surface density and drive vortex formation,
the smooth capture of planets into mean motion resonances is disrupted.
As planets pile up in close orbits, not protected by resonances, close encounters increase the probability of planet-planet collisions,
even while the gas disc is still present.
While inviscid discs often produce unstable non-resonant systems, stable, closely packed, 
non-resonant systems can also be formed.
Thus, when examining the expectation for planet migration to produce planetary systems in mean motion resonances, 
the effective turbulent viscosity of the protoplanetary disc is a key parameter.
 \end{abstract}




\section{Introduction}

Convergent migration in multiplanet systems, driven by disc-planet interactions in protoplanetary discs, has been shown to result in the capture of the planets into mean motion resonances \citep[hereafter MMRs,][]{2001A&A...374.1092S, 2002ApJ...567..596L,2002MNRAS.333L..26N}.
\citet{2006A&A...450..833C} tested the behaviour of initially tightly-packed systems in viscous discs, and 
found that after a period of initial adjustment almost all systems formed chains of MMRs.
However, the multiplanet systems discovered by the Kepler mission have only a weak preference for period 
ratios near first order mean motion resonances \citep{2011ApJS..197....8L}.
Multiplanet systems in the Kepler sample tend to have planets with similar masses, 
with relatively even orbital spacing \citep{2017ApJ...849L..33M,2018AJ....155...48W}, 
but are largely non-resonant. Furthermore, the sample contains planets which appear to mainly cluster around the local thermal mass scale in a fiducial protoplanetary disc model \citep{2018arXiv180604693W}, where the thermal mass corresponds to the planet Hill sphere radius being approximately equal to the disc pressure scale height.

Various mechanisms have been proposed to either allow planets to escape a resonant configuration during the presence of the gas disc, 
or to disrupt the resonant configuration during the later, nearly dissipationless, n-body phase of dynamics.
Overstable librations about resonant configurations can cause planets to escape resonance while the gas disc is present \citep{2014AJ....147...32G}, although the requirements on the form of the eccentricity damping for this to occur may be difficult to meet.
Disc-driven resonant repulsion can push a resonant pair of planets away from resonance by a combination of orbital circularization and the interaction between the wakes of the planets \citep{2013ApJ...778....7B}.
Orbital perturbations due to turbulent overdensities in the disc have been argued to prevent protoplanets
  capturing into resonance \citep[e.g.][]{2008ApJ...683.1117A}. 
However, the planet forming regions of protoplanetary discs are thought to be largely dead to the magnetorotational
  instability (MRI), and lacking an instability capable of driving turbulent motion at such high levels.
If protoplanetary discs are characterised in these regions by being largely MRI-dead, possessing nearly-laminar flow with wind-driven accretion in their surface layers,  
they can possess vanishingly low viscosity while still providing a conduit for mass accreting onto the star \citep{2013ApJ...769...76B}.

A second set of concepts proposes that a resonant chain of planets
may be disrupted after dissipation of the gas disc. \cite{2007ApJ...654.1110T} showed that tidal interaction with the central star could extract short period systems out of resonance.
\cite{2015ApJ...803...33C} proposed the interaction of the planets in resonant orbits with a 
planetesimal disc  left over from planet formation may break planets out of resonance.
More ambitiously,
the possibility that late dynamical instability of resonant chains formed through convergent 
migration in a viscous disc is responsible for sculpting the entire period ratio distribution of exoplanet systems was raised by \citet{2014A&A...569A..56C}.
\citet{2017MNRAS.470.1750I} and \citet{2019arXiv190208772I} demonstrate that 
systems with large numbers of protoplanetary cores in an n-body computation with a prescription for disc-planet interactions 
may result in a planetary system configuration which becomes unstable after the dissipation of the gas disc.
This behaviour appears to be due to the increasing tendency for chains with a high mass to undergo dynamical instability after the dissipation of the gas disc, 
as identified by \citet{2012Icar..221..624M}.

In this letter, motivated by our recent work showing that disc-planet interactions involving intermediate mass planets embedded in inviscid protoplanetary discs leads to stochastic, non-deterministic migration behaviour due to the emergence of vortices in the flow \citep{2019MNRAS.484..728M}, we question the basic premise that convergent migration in 
 protoplanetary gas discs should result  in chains of planets in mean-motion resonance,
and the consequent tendency to form systems of resonant planets which are stable over Gyr time scales.
We construct a scenario where a like-for-like comparison between the convergent migration 
of a multiplanet system in a viscous and an inviscid disc can be made, and demonstrate that, in contrast to the situation in viscous discs, 
the ability to form resonant chains of planets is impeded by vortex-modified feedback migration in inviscid discs.

\section{Methods and Results}

\begin{figure*}
\begin{center}
\includegraphics[width=1.8\columnwidth,trim=0 0.1cm 0 0, clip]{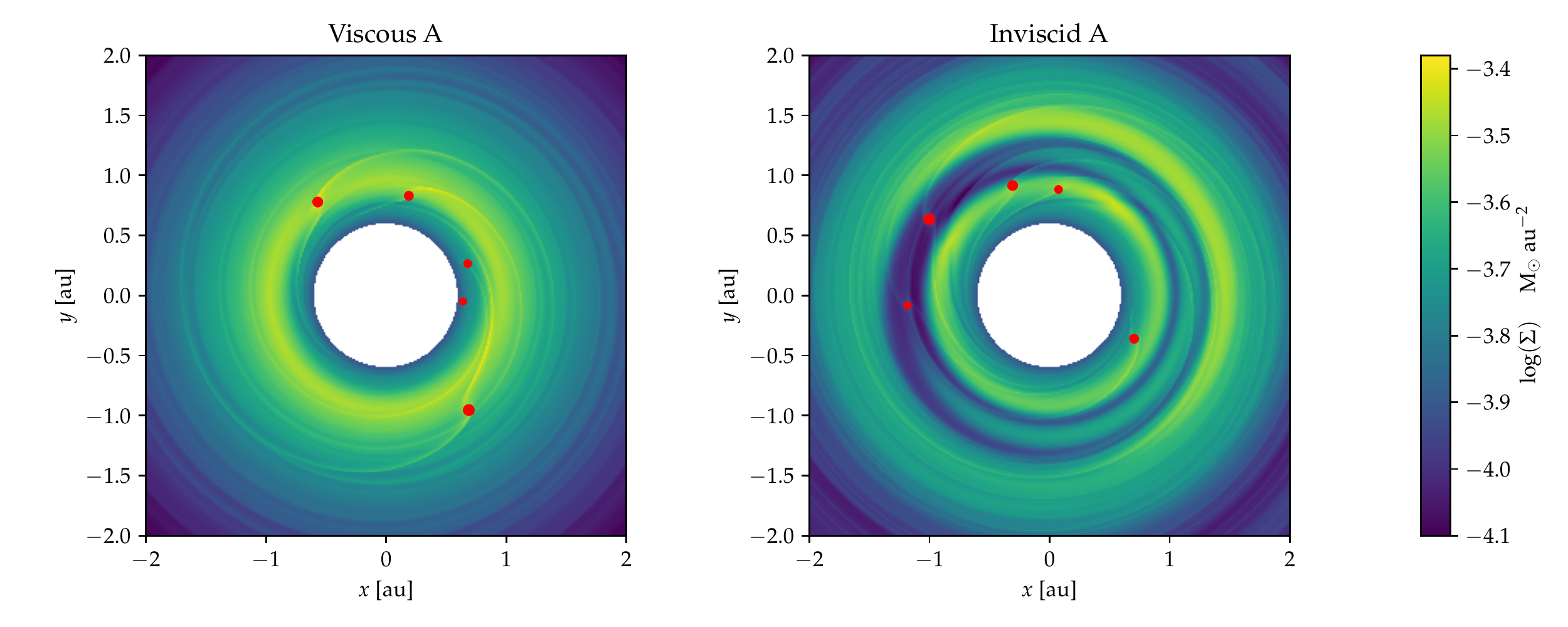}
\end{center}
\caption{Disc gas surface density, with planet positions in example viscous and inviscid discs after 60000 years. 
Red circles indicate planet positions, with symbols indicating their relative mass. 
The viscous case produces a smooth disc with a spiral wake from each planet,
while the inviscid case produced additional vortices and a significant partial gap dug by the most massive planet.}
\label{fig:logsigma comparison}
\end{figure*}

\begin{figure}
\begin{center}
\includegraphics[width=\columnwidth]{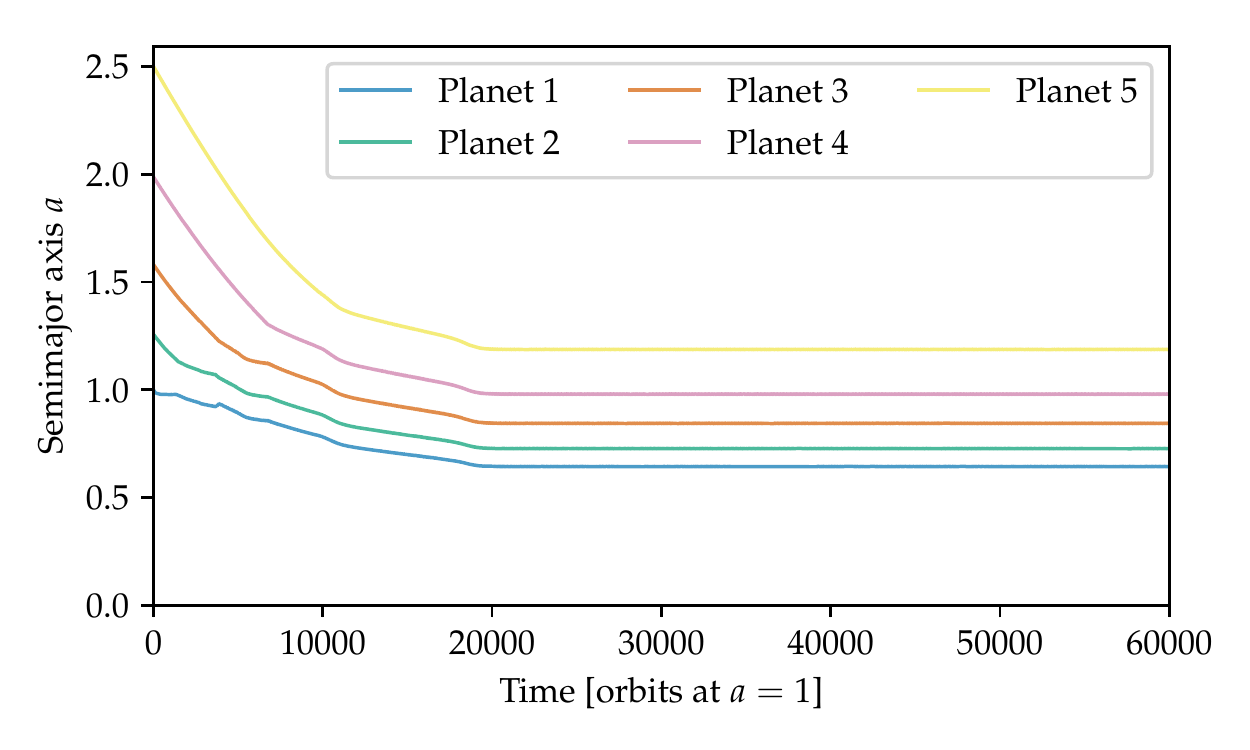}
\end{center}
\caption{Planet migration in a viscous disc, compare to Figure~\ref{fig:convsys29a}.
All five planets form a resonant chain, and in long term evolution this configuration tends to be stable.}
\label{fig:convsys30a}
\end{figure}

\begin{figure}
\begin{center}
\includegraphics[width=\columnwidth]{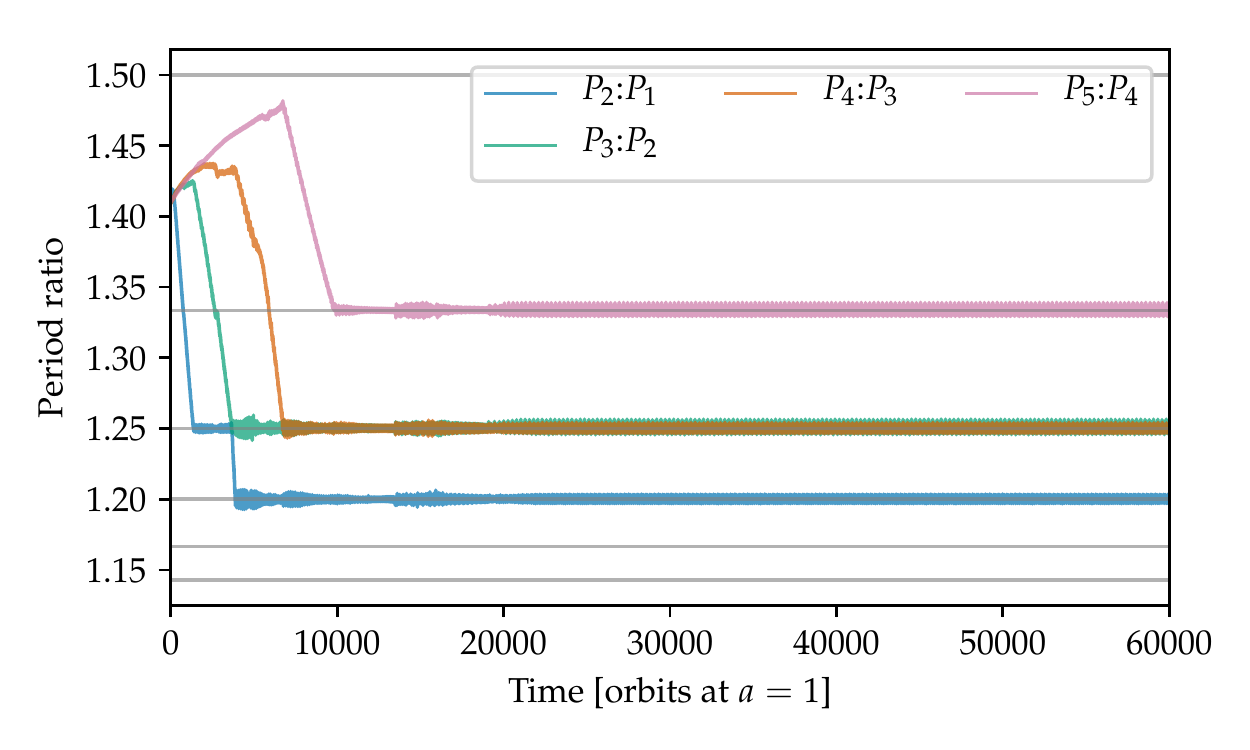}
\end{center}
\caption{Planet nearest neighbour period ratios in a viscous disc, displaying the formation of a resonant chain.}
\label{fig:convsys30pratio}
\end{figure}

\begin{figure}
\begin{center}
\includegraphics[width=\columnwidth]{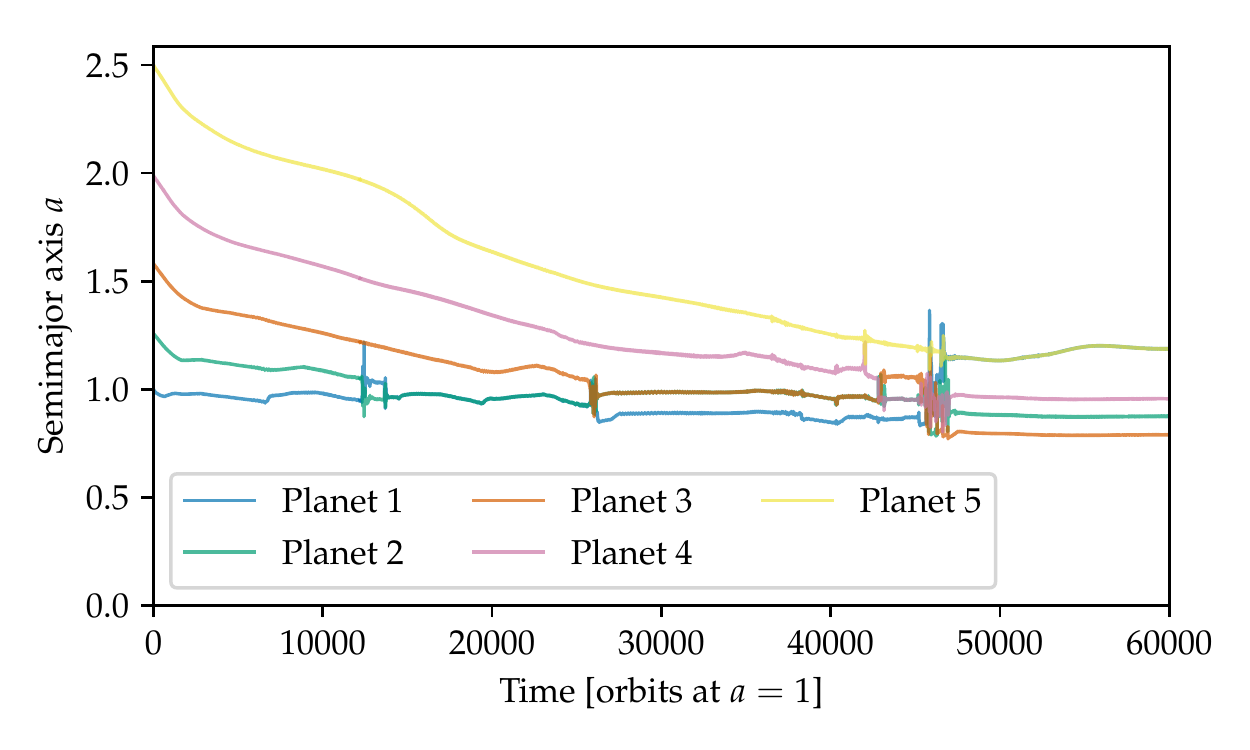}
\end{center}
\caption{Planets in an inviscid disc, compare to Figure~\ref{fig:convsys30a}. The final configuration consists of three planets that have migrated into the viscous region of the disc in a chain of resonances, and a coorbital pair to the outside in the inviscid region out of resonance. Sampling the long term behaviour of this system shows it tends to undergo dynamical instability.}
\label{fig:convsys29a}
\end{figure}

Gas disc-planet interaction simulations were performed in two-dimensional vertically integrated models of gas discs with a 
modified version of {\sc FARGO3D}~1.2 \citep{2016ApJS..223...11B}, including an implementation of the energy equation in term of 
specific entropy (see Appendix~\ref{sec:entropy}).
Indirect terms for the planets and gas disc were included, the planet potential was smoothed with a Plummer-sphere potential with length 0.4 scale heights,
 and the disc gravity force felt by the planets was modified by removing the 
azimuthally symmetric component of the potential to compensate for the neglect of disc self-gravity following \citet{2008ApJ...678..483B}. 
In {\sc FARGO3D} the planet orbits are integrated with the built-in 5th order scheme, 
with an additional planet-planet acceleration time step limit from \citet{2006A&A...450..833C} to increase accuracy  of energy conservation during close encounters.
The detailed outcomes of these close encounters and three-body interactions is chaotic and senstive to small perturbations in 
the initial conditions and the numerical method of integration.
We do not include planet-planet collisions.
The grid spacing was radially logarithmic, extending radially from $r=0.6$ to $4$, with resolution corresponding to $\sim 24$ cells per scale height in all directions. Damping boundary zones were applied as in \citet{2019MNRAS.484..728M}, 
The azimuthal velocity field was initialised to produce an exact numerical radial force balance, following the method 
implemented in {\sc FARGO} \citep{2000A&AS..141..165M,2011ascl.soft02017M}.
The runs presented in this work required in total 600 kCPUh.

Disc thermodynamics were modelled in the simplest useful form for considering the near-adiabatic 
thermodynamics of the inner regions of protoplanetary discs in two dimensions.
Thus, we apply a thermal relaxation term 
in the form used by \citet{2012ApJ...750...34H} and \citet{2016ApJ...817..102L}, with a timescale derived from an effective optical depth estimate 
from \citet{1990ApJ...351..632H} for an irradiated disc as described by \citet{2012ApJ...757...50D}.
We adopt the simplified Rosseland mean opacity model of \citet{1997ApJ...486..372B} and 
approximate the Planck mean opacity as being the equal to it. 
In viscous (turbulent) disc regions, we include a subgrid turbulent diffusivity to entropy assuming a turbulent Prandtl number unity, i.e.~equal diffusion of momentum and specific entropy.
One consequence of including the energy equation in this manner is that vortices can form due to both the Rossby Wave Instability (RWI) and through baroclinic forcing. This is in contrast to our recent work where a barotropic equation of state was adopted \citep{2019MNRAS.484..728M}. In terms of realism, the thermodynamic treatment adopted in this paper is an improvement on our previous approach.

To facilitate a comparison on as equal terms as possible between the behaviours of viscous and inviscid discs, we have constructed a model for a disc with 
a planet migration trap formed at the inner radial edge of the dead zone where the disc transitions from being MRI inactive to becoming turbulent.
This radial location is a density maximum, with a rapidly decreasing surface density to the inside where the disc has a large viscosity. Exterior to this edge the disc is inviscid or has a lower viscosity.
Thus we must arrange for a disc which has a 
stationary configuration under the action of viscosity even though it has
1) a non power-law radial surface density profile, and
2) it may have a smooth transition from viscous to inviscid.
These two properties can be arranged by modifying the viscosity operator to diffuse the disc towards the initial surface density profile.
This is implemented by subtracting a term equal to the initial specific viscous torque from the azimuthal momentum equation, supplementing equation~(129) of \citet{2016ApJS..223...11B} with
\begin{align}
\partial_t  v_{\phi} = \ldots - \left[ - \frac{1}{\rho}\left\{ \frac{1}{r^2}\partial_r (r^2 \tau_{\phi r}) \right\} \right]_{t=0}\ .
\end{align}
In the case of an inviscid outer disc region, $\tau_{\phi r}=0$ and the additional viscous force is zero.
With this added term, the initial condition density profile is an equilibrium state for the disc.

The disc surface density profile is $\Sigma = \Sigma_0 (r/0.8)^{5/2}$ inside of $r=0.64$ and 
$\Sigma = \Sigma_0 r^{-3/2}$ outside of $r=0.84$ and 
given by the the unique smooth cubic interpolating polynomial in the interval $r=[0.64,0.84]$.
The disc has an initial radial temperature distribution $T\propto r^{-3/7}$ and aspect ratio
$h= 0.035 r^{2/7}$ corresponding to a passively irradiated disc.
When used, the viscosity is scaled radially to produce no net accretion flow in the two power-law 
regimes of the disc, so $\nu=\nu_0 r^{\alpha-1/2}$ with $\alpha=-5/2$ for $r<0.8$ and  $\alpha=3/2$ for $r>0.8$. 
In the viscous case, the viscosity scaling coefficient is everywhere $\nu_0=2\times10^{-5}$.
At r=1, this viscosity corresponds to a Shakura-Sunyaev turbulent viscosity of $\alpha_{\rm SS}=8\times10^{-3}$.
In the case representing a laminar dead zone the viscosity coefficient $\nu_0$ is tapered 
linearly to zero in the interval $r=[0.85,0.95]$ and is zero at larger radii.
The planetary system is characterised by a factor of $5/2$ in planet mass 
($q=\left[ 1, 1.26, 1.58, 1.99, 2.5 \right] \times 10^{-5}$) and in $5/2$ in initial orbital semimajor axis ($a=1, 1.26, 1.58, 1.99, 2.5$).
The planet masses thus range from approximately 3 to 8 solar masses, making them similar to the Kepler multiplanet sample.
As these planets migrate inwards, they are between the local feedback mass and thermal mass, where at the feedback mass spiral density waves excited by a planet undergo quasi-local shock dissipation resulting in partial gap formation \citep{2002ApJ...572..566R}.

In the globally viscous disc case (left panel of Figure~\ref{fig:logsigma comparison}), the planets settle into a chain of first order mean motion resonances, 
and the planets stay well separated in their orbits at all times, as shown in Figure~\ref{fig:convsys30a} and Figure~\ref{fig:convsys30pratio}.
However, in the inviscid disc (right panel of Figure~\ref{fig:logsigma comparison}), where the planets drive significant structure in the disc, including vortices, 
some planets are able to escape mean motion resonances, and undergo a series of chaotic close encounters as shown in Figure~\ref{fig:convsys29a}.
Data for a second viscous disc system and three more inviscid discs systems with perturbed initial planet positions, 
 showing this behaviour is a generic outcome, are presented in  the Supplementary Material (online only).

Once these compact systems have been produced, either in a chain of resonances or not, the further evolution of the system
in the pure n-body post-disc phase can be examined.
To model this, we restart each run at 11 times, evenly spaced from $45\times10^5$ to $60\times10^5$ orbits, and 
add a operator to the surface density equation $\partial_t \Sigma = -1/t_e \Sigma$ 
with the disc evaporation timescale $t_e=800$~orbits.
This timescale is chosen to be as short as reasonable which maintaining slow, adiabatic changes to the n-body system.
After 5 e-folding times of this disc evaporation operator, the planet and star positions and velocity configuration were saved for use in the n-body phase.
This  pure n-body evolution was computed with {\sc REBOUND} \citep{2011ascl.soft10016R,2012A&A...537A.128R} 
using the WHFast integrator \citep{2015MNRAS.452..376R}
with a time step of $1/120$ of an orbital time at $a=1$, continued until a maximum time of  $10^8$ orbits was reached, 
or when any individual planet achieved an eccentricity above $0.95$.
Thus close encounters are not directly detected or given special treatment, as no attempt is made to integrate past their occurrence in this post-gas-disc phase.
This eccentricity stopping criteria indicates that the system likely undergoes close encounters or dynamical instability.
In total, the viscous disc systems survive to $10^8$ orbits in a chain of mean-motion resonances, with 5 exceptions out of 22 cases.
However, of the 44 cases of the inviscid discs, only 11 survive for $10^8$ orbits, with 7 of those
occurring for n-body extractions from a single gas disc evolution case. 
It is important to stress that while these results cannot be interpreted as a general statistical statement for the frequency of stable systems, 
they demonstrate both that the systems resulting from migration in an inviscid disc are in the long term much more likely to undergo dynamical instability
as the result of the planets not being entirely locked in a chain of first-order mean-motion resonances,
and additionally that the inviscid disc cases can produce systems with planets out of resonance that are nevertheless stable for long periods. 
The semimajor axis and eccentricity evolutions of the 55 realisations of full system histories are presented in Supplementary Material (online only),
along with the histories of the nearest first-order mean-motion resonant angles.

\section{Discussion}
The stable systems resulting from viscous disc migration were always found to be in mean motion resonance, 
but those resulting from migration in an inviscid disc may not form complete resonant chains, despite being closely packed.
This phenomenon may provide a route to forming systems like Kepler-11, which contain a chain of closely-packed planets close to, 
 but not in mean motion resonances \citep{2014ApJ...795...32M}.
 In such a scenario the formation of the planets at precisely their late-time orbital locations, 
 and the suppression of any planet-disc interactions leading to planet migration \citep[in-situ formation,][]{2012ApJ...751..158H} is not required. Furthermore, the inviscid disc scenario provides an attractive alternative to the idea that Kepler-11 may have previous been in a resonant chain that subsequently broke up due to dynamical instability, as envisioned in the recent calculations of \citet{2019arXiv190208772I}. Systems displaying the close-packed nature of Kepler-11, with its proximity to dynamically unstable configurations, do not arise easily from scenarios that involve strong scattering and dynamical relaxation of multiplanet systems \citep{2014ApJ...795...32M}, and hence a gentler formation scenario involving migration in an inviscid disc may explain these types of systems.

While our simulations include a fairly complete physical model, numerous improvements will be required in future work to confirm the results we have presented here.
The thermal relaxation model adopted has been sufficient for this study, 
however the heating of the disc due to the tidal interaction of feedback-mass 
planets can be significant over the long timescales considered here.
More accurate predictions of the outcome of low-viscosity planet-disc interactions will 
require a radiative transfer prescription valid far from the disc irradiation-cooling equilibrium.

Our models have constrained the planet orbits to zero inclination. Planet-planet and planet-disc interactions
 will have an effect on the evolution of this parameter. Although allowing non-zero inclination would not be expected to change the 
qualitative result about the difference between system evolution in viscous and inviscid discs found here, 
the possible quantitative differences in the results of systems produced from viscous and inviscid discs and comparison to observations is an important area for follow up work.

In inviscid discs, many close encounters between pairs of planets occur when the gas is still present, 
in strong contrast to the viscous disc models, where efficient capture into resonance prevents close planet-planet encounters.
We note that these close encounters often lead to separations which under reasonable planet mass-radius models
a collision would be inevitable, if this process were included in our model.
However, as the planet trajectories were constrained to zero inclination the collision probability 
during a close encounter is significantly increased as compared to true three dimensional interactions.
Further work should include a three dimensional treatment of the planet trajectories, and a model for planet-planet collisions as these may become much more important for the evolution of planetary systems in inviscid discs.
Dynamical instability in the post-gas disc should drive the loss of planet atmospheres through heating \citep{2019MNRAS.tmp..739B},
but collisions while the gas disc is still present may allow the planet to re-accrete a gas atmosphere.

\section{Conclusions}
Convergent migration has different typical outcomes in viscous and inviscid discs.
In viscous disc models, planets are able to migrate into resonant chains, with a 
preference for first-order mean motion resonances.
In inviscid discs, the ability of planets, particularly those above the feedback mass, 
to spur vortices and modify the disc surface density profile 
both allows planets to escape resonant configurations and to migrate into 
sustained non-resonant configurations.
These non-resonant configurations have significantly different long-term stability 
properties from chains of planets in mean motion resonances, being much  more likely to be unstable.
At the same time, the ability of planets to escape mean-motion resonances while undergoing convergent migration in an inviscid disc allows the
system to find tightly packed configurations which appear to possess long term stability without being entirely in mean motion resonances.

\section*{Acknowledgements}
This research was supported by STFC Consolidated grants awarded to the QMUL Astronomy Unit 2015-2018  ST/M001202/1 and 2017-2020 ST/P000592/1.
This research utilised Queen Mary's Apocrita HPC facility, supported by QMUL Research-IT \citep{apocrita};
and the DiRAC Data Centric system at Durham University, operated by the Institute for Computational Cosmology on behalf of the STFC DiRAC HPC Facility (www.dirac.ac.uk).
We acknowledge PRACE for awarding access to TGCC Irene at CEA, France.
This equipment was funded by a BIS National E-infrastructure capital grant ST/K00042X/1, STFC capital grant ST/K00087X/1, DiRAC Operations grant ST/K003267/1 and Durham University. 
DiRAC is part of the National E-Infrastructure.
SJP is supported by a Royal Society University Research Fellowship.
Simulations in this paper made use of the REBOUND code which can be downloaded freely at {\tt http://github.com/hannorein/rebound}.




\bibliographystyle{mnras}
\bibliography{resplanets_paper} 



\appendix
\section{Specific entropy formulation}
\label{sec:entropy}

FARGO3D by default evolves energy by use of an 
internal energy variable (internal energy per volume).
Like other ZEUS-family codes \citep{2011IAUS..270....7N}, the variable used to evolve energy can be changed.
For example, in ZEUS, the use of a total energy formulation has been shown to have important advantages \citep{2010ApJS..187..119C}.  The specific algorithmic choices made in updating the energy variable are also not canonical, and variations to the use of ``consistent advection'' \citep{1980ApJ...239..968N} may be advantageous for the energy variable \citep{2010ApJS..187..119C}.


In this work, we have opted to evolve the specific entropy, for the reason that like 
the internal energy, the evolution requires only the use of cell-centred quantities, because the evolution equation does not 
have a compression-work term.
Specific entropy $s$ is the entropy per mass. 
With the ideal gas equation of state as:
\begin{align}
P  &= (\gamma-1)\Sigma c_v T\ ,
\end{align}
where $P$ is the gas pressure, $\gamma$ the adiabatic index, $\Sigma$ the gas surface density, and $T$ the temperature,
 the heat capacity at constant volume is defined as:
 \begin{align}
 c_v \equiv \frac{k_B}{\mu m_H (\gamma-1)}\ ,
 \end{align}
 with $k_B$ the Boltzmann constant, and $\mu m_H $ the mean mass per gas particle.
 The specific entropy itself is thus defined as:
 \begin{align}
s\equiv c_v \log\left(\frac{T}{T_0}\left(\frac{\Sigma}{\Sigma_0}\right)^{-(\gamma-1)}\right)\ ,
\end{align}
where $\Sigma_0$ and $T_0$ are constant reference densities and temperatures, and play not role in determining the gas physics.
 
The evolution equation for specific entropy is
\begin{align}
\frac{\partial s}{\partial t} &= - (\V \cdot \nabla) s + \mathcal{L} \label{eq:sevol}\ ,\\
\mathcal{L}&\equiv \frac{1}{T}\left( \Gamma_\nu +\Gamma_{\rm sh}\right)  + \frac{1}{T}\left[ -c_v \frac{T-T_{\rm ref}(r)}{t_{\rm cool}}\right]
\end{align}
where $\mathcal{L}$ is the heating/cooling term, containing the entropy production 
in viscous dissipation (both shock-capturing artificial viscosity and kinematic viscosity) and a thermal relaxation term \citep{2016ApJ...817..102L}.
The viscous heating $\Gamma_\nu$ and shock-capturing artificial viscosity heating $\Gamma_{\rm sh}$ remain the same as in FARGO3D's implementation of the internal energy equation.
The motivation for the choice of this form of energy equation is that it lacks a PdV term that the internal energy density evolution equation has.
ZEUS-family schemes such as FARGO3D utilise a simple form of operator splitting to solve the fluid evolution equations, 
where the fields are updated by a full time step in the source terms, and then the fields are 
updated by the conservation law component of the evolution equation.
Thus, to integrate equation~(\ref{eq:sevol}) only the source term $\mathcal{L}$ is integrated during the source step.
The remainder of equation~(\ref{eq:sevol}) is applied during the transport step.

The transport operator component of equation~(\ref{eq:sevol}) is
\begin{align}
\frac{\partial s}{\partial t} &= - (\V \cdot \nabla) s\ .
\end{align}
To solve this equation in FARGO3D we multiply $s$ by density $\rho$ before the transport step, and so solve for the transport of the volumetric quantity $S=\rho s$.
The quantity $S$ obeys a conservation law as 
\begin{align}
\frac{\partial S}{\partial t} + \nabla \cdot (S\V) = 0 \ .
\end{align}
After the update of $S$ to the end of the time step, the result is transformed back to the specific entropy by utilising the updated density field. 
The series of steps used to apply the  transport operator from time $t_i$ to $t_{i+1}$ to the density and entropy fields  in the transport step is thus
\begin{align}
S_{i} &= s_{i} \rho_{i}\\
s_{i+1}& = \mathrm{Transport}(s_i)\\
\rho_{i+1} &= \mathrm{Transport}(\rho_i)\\
s_{i+1} &= S_{i+1}/\rho_{i+1}
\end{align}
Lacking a strong reason to adopt ``consistent advection'' on top of the procedure of transforming to $S$ 
with the initial density, and back to $s$ with updated density, we follow the advice 
given in \citet{2010ApJS..187..119C} and disable it for the transport of $S$.
As a validation of the implementation, a version of the Sod shock tube test is shown in the Supplementary Material.

\bsp	
\label{lastpage}
\onecolumn
\pagebreak


\section*{Supplementary material}

# Multiplanet systems in inviscid discs can avoid forming resonant chains: Extended online appendices

Colin P. McNally,[1]* Richard P. Nelson[1] and Sijme-Jan Paardekooper [1,2]
[1]Astronomy Unit, School of Physics and Astronomy,
Queen Mary University of London, London E1 4NS, UK
[2]DAMTP, University of Cambridge, Wilberforce Road, Cambridge CB3 0WA, UK

## B  Online Appendix: Extended data on all gas disc runs

Four inviscid disc systems, labelled Inviscid A-D and two viscous disc systems labelled Viscous A-B were simulated. Figure 1 presents the evolution of 4 systems in an inviscid disc, with the latter three runs having initial semi-major axis of the planets increased by 1%, 2%, and 3% over the first respectively. Figure 2 presents the evolution of 2 systems in a viscous disc, with the latter run having initial semi-major axis of the planets increased by 1% over the first.

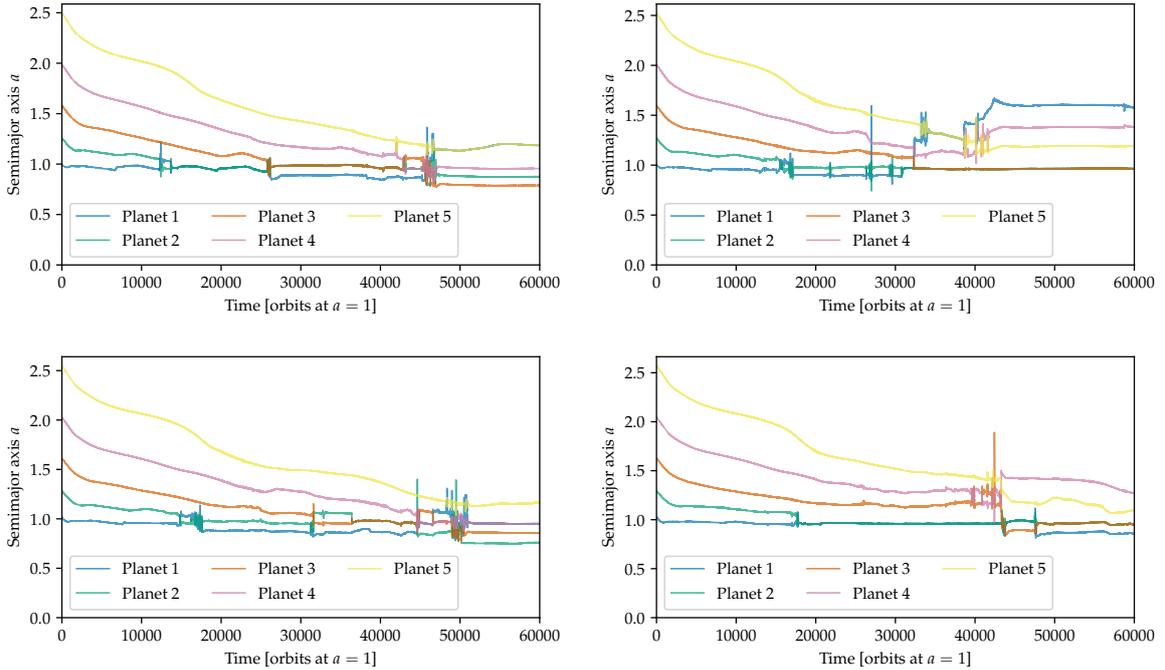

Figure 1: Planet migration in a inviscid disc with varying initial semi-major axes, upper left panel repeats that shown in main article text. Upper row: Inviscid A, Inviscid B, Lower row: Inviscid C, Inviscid D.

---

*E-mail: c.mcnally@qmul.ac.uk (CPM)

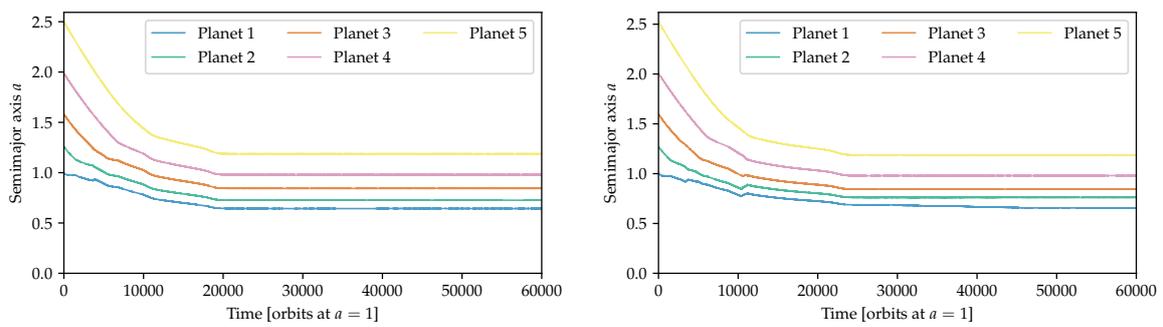

Figure 2: Planet migration in a inviscid disc with varying initial semi-major axes, left panel repeats that shown in main article text. Left: Viscous A Right: Viscous B

# C  Online Appendix: All system histories, gas disc and n-body phases

## C.1  Orbital semimajor axis for all realizations

These figures show the orbital semimajor axes for the entire history of each system computed. For each gas disc simulation, 11 times are selected for the beginning of gas disc evaporation, and after the gas disc has been removed the system is evolved in a pure n-body member (see Methods & Results). In these plots, the evolution during the gas disc, gas disc evaporation, and n-body phase is combined and shown as a full time evolution. The four inviscid gas disc runs and two viscous gas disc runs are each presented on a panel with 11 rows, one for each disc evaporation time.

## C.2  Orbital eccentricity for all realizations

Following the same formatting as the previous section, these figures show the orbital eccentricity for the entire history of each system computed.

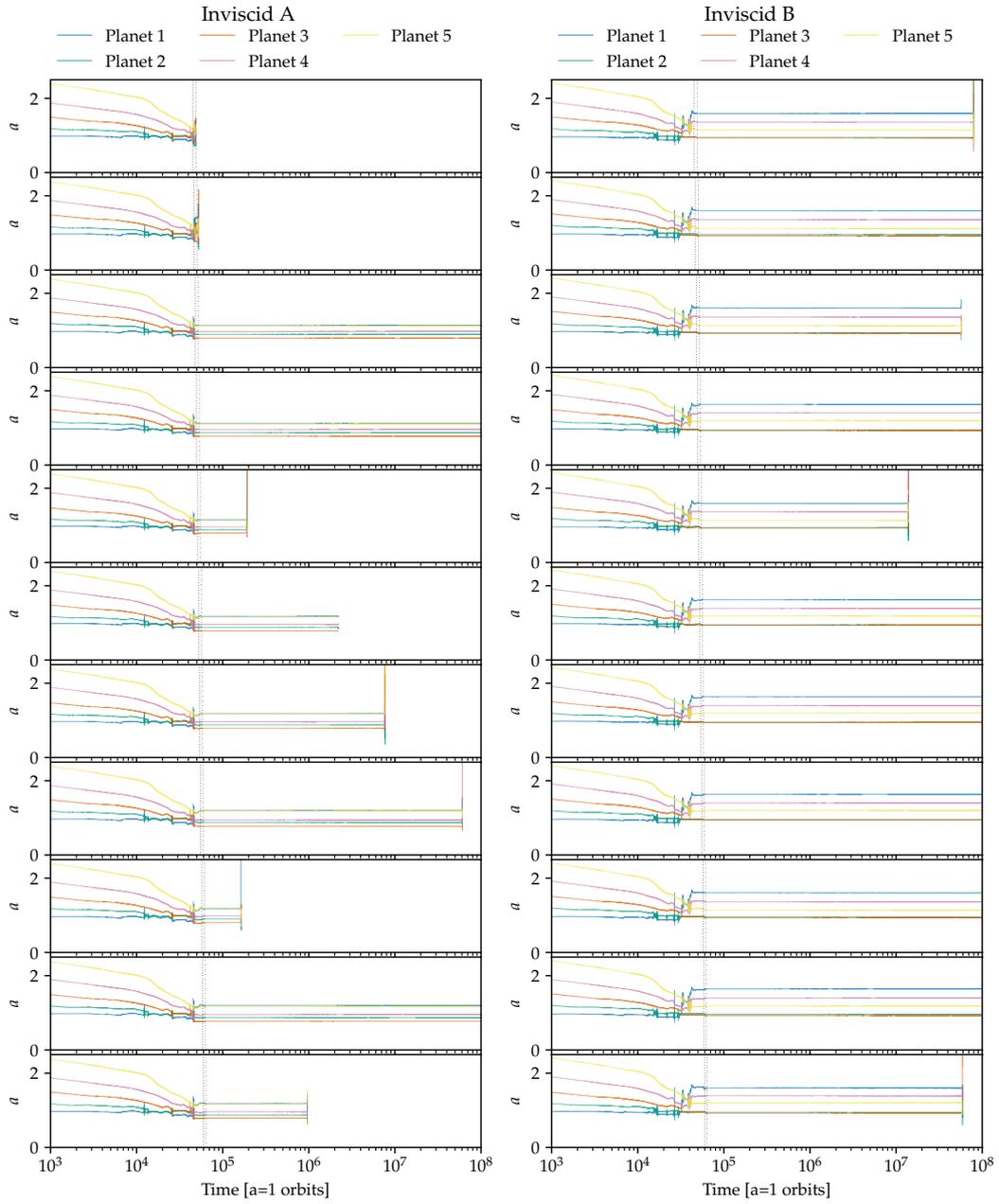

Figure 3: Planet migration in a inviscid disc with varying initial semi-major axes including disc dissipation and pure n-body phase.

4

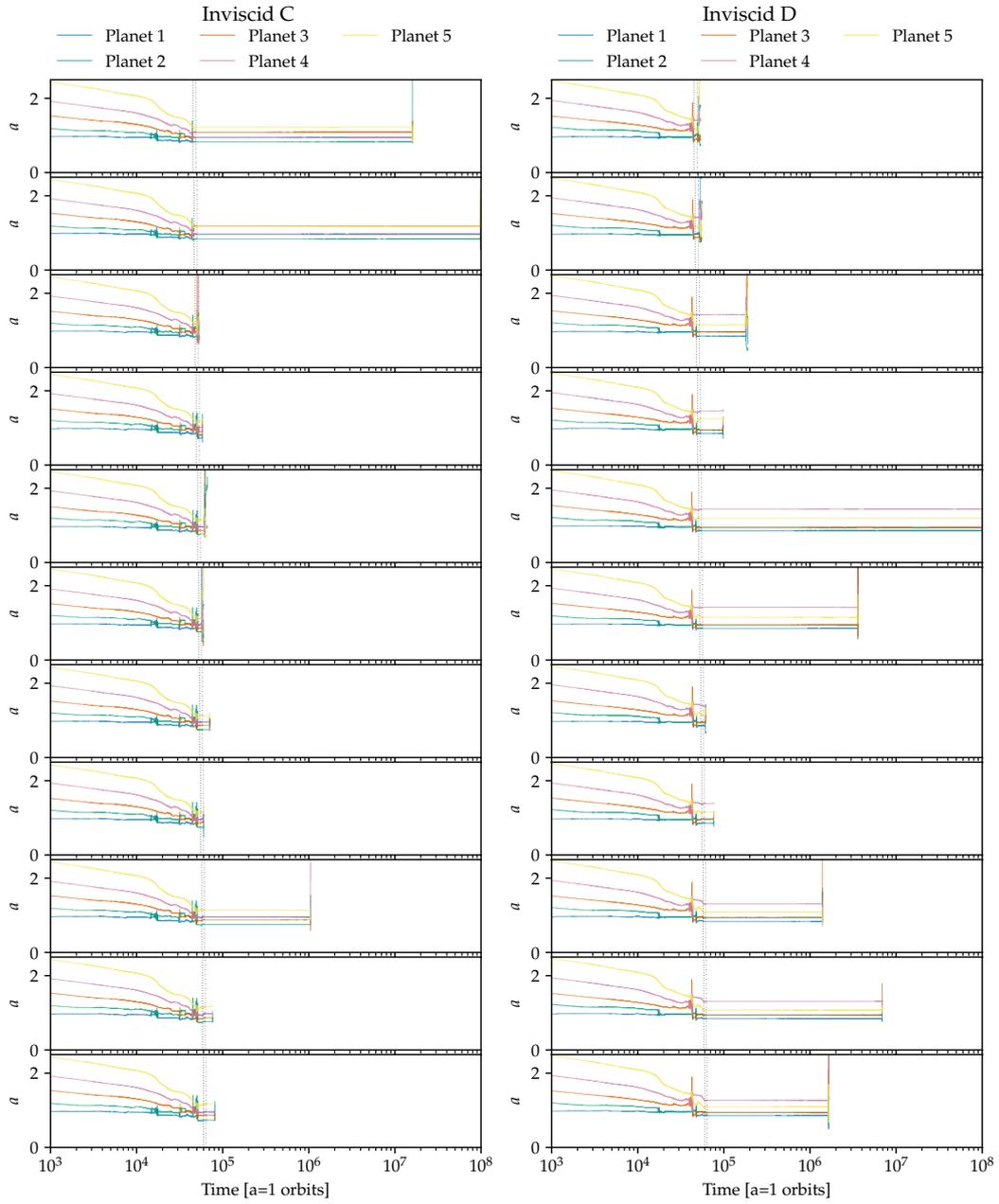

Figure 4: Planet migration in a inviscid disc with varying initial semi-major axes including disc dissipation and pure n-body phase.

5

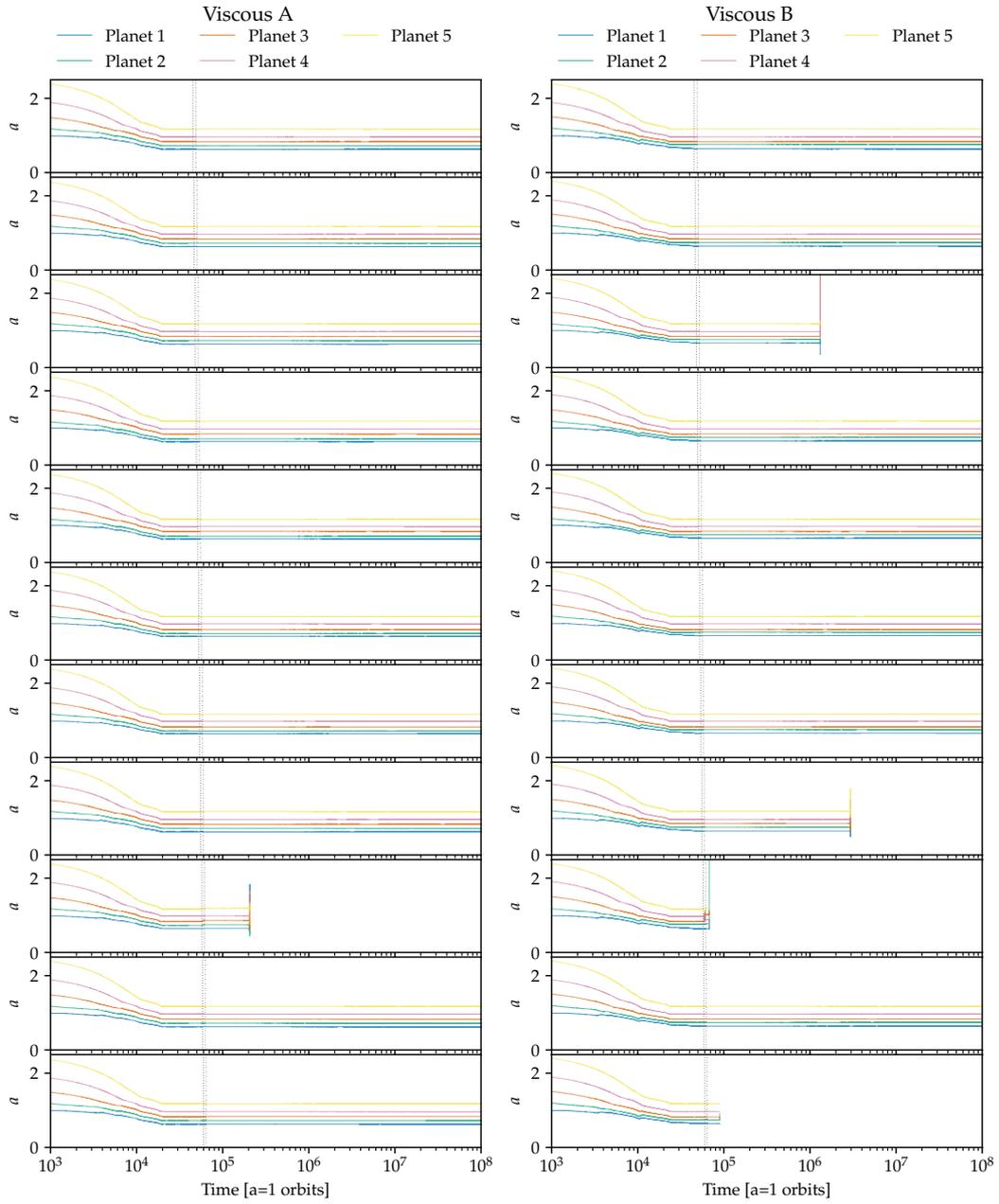

Figure 5: Planet migration in a viscous disc with varying initial semi-major axes including disc dissipation and pure n-body phase.

6

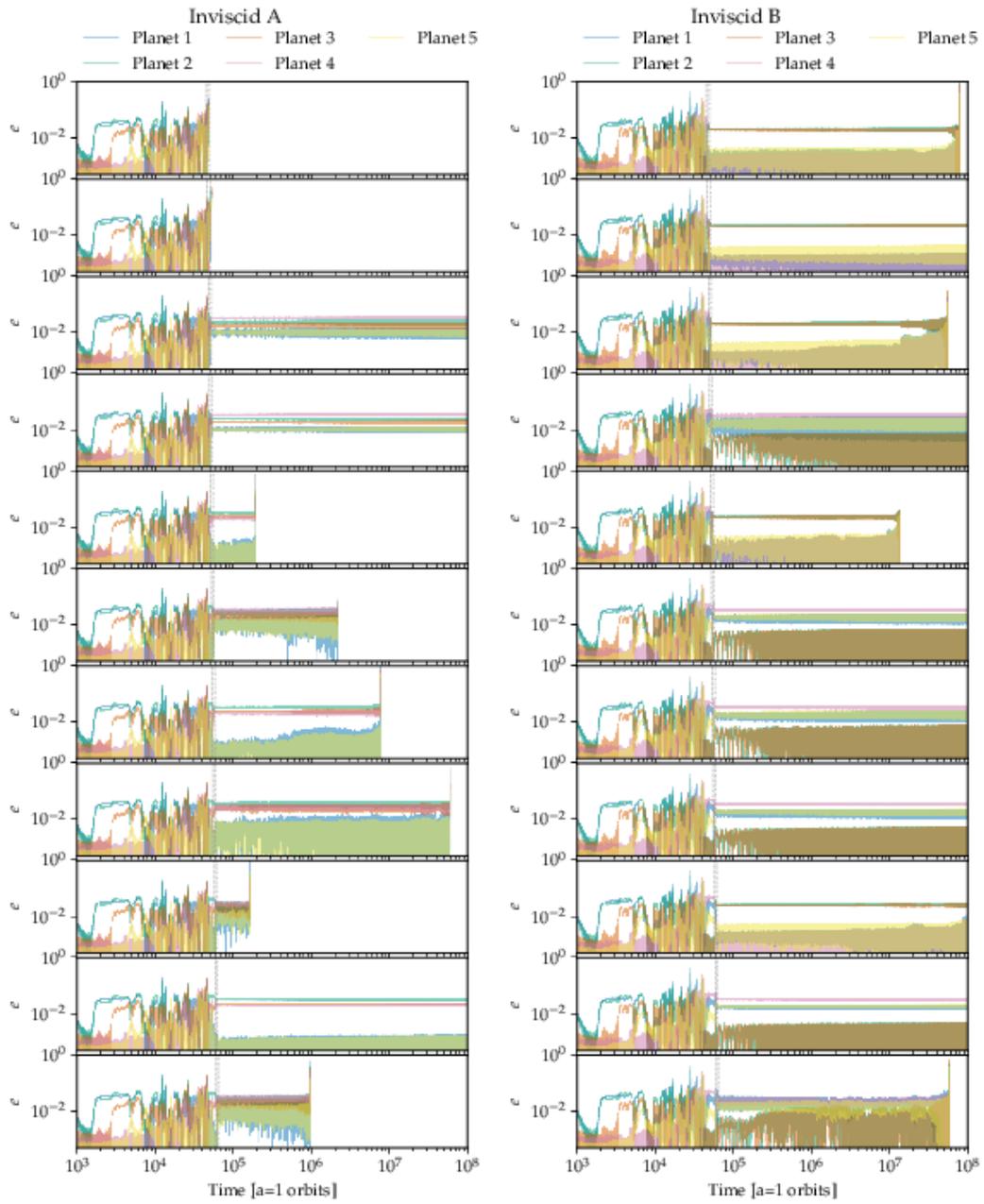

Figure 6: Eccentricity history in a inviscid disc with varying initial semi-major axes including disc dissipation and pure n-body phase.

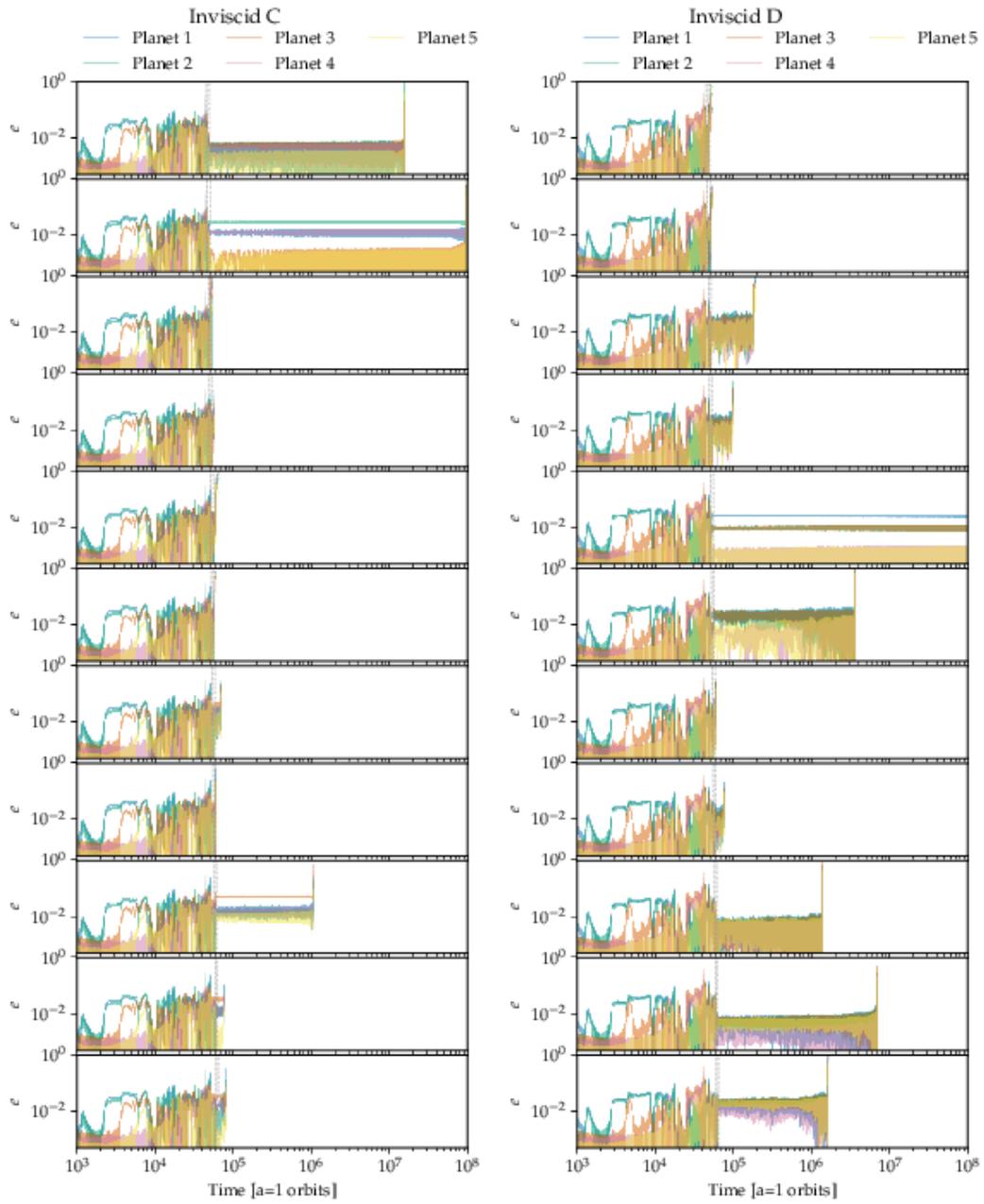

Figure 7: Eccentricity history in a inviscid disc with varying initial semi-major axes including disc dissipation and pure n-body phase.

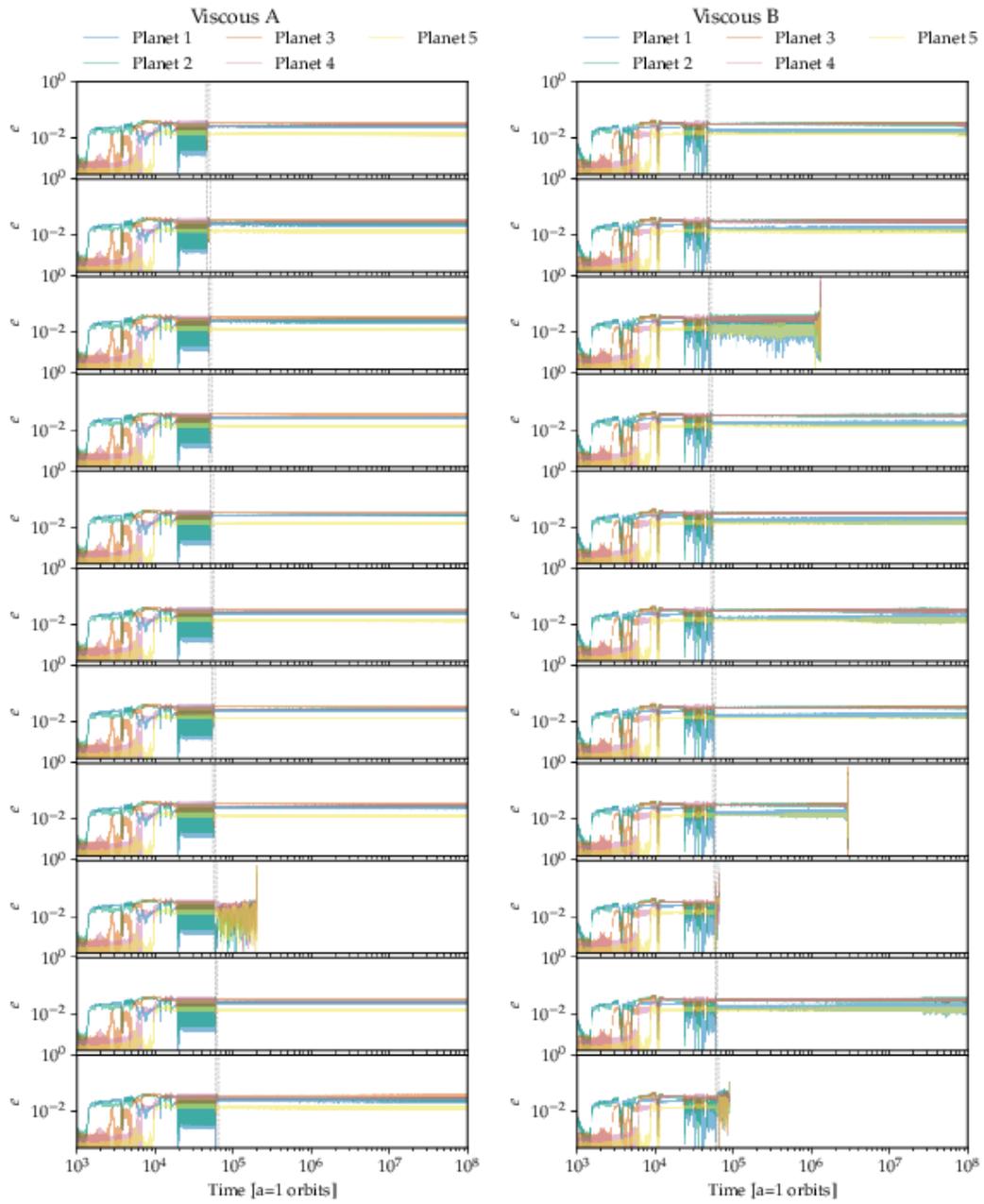

Figure 8: Eccentricity history in a viscous disc with varying initial semi-major axes including disc dissipation and pure n-body phase.

## C.3 Resonant angles for late-time configurations

These figures resent the resonant angles for the full system lifetime, for the nearest first-order mean-motion resonance identified for each nearest pair of planets in semimajor axis at late times. They identify when planets in a chain are captured in a mean-motion resonance. Each panel is identified by the parent gas disc run, and labelled in order by the time it was extracted to the gas disc evaporation phase, corresponding the row number in the plots of the previous section. For a first order resonance $p{:}q$ (i.e. $q + 1{:}q$) the two resonant angles are given by:

$$\phi_1 = p\lambda_1 - q\lambda_2 - \omega_1 \tag{1}$$
$$\phi_2 = p\lambda_1 - q\lambda_2 - \omega_2 \tag{2}$$

where $\lambda_1$ is the mean longitude and $\omega_1$ the longitude of pericentre (equivalently in this case periapsis, perifocus, or periastron) of the outer planet, and $\lambda_2$, $\omega_2$ the same for the inner planet. Planets in coorbital paris are identified as being in a 1:1 resonance.

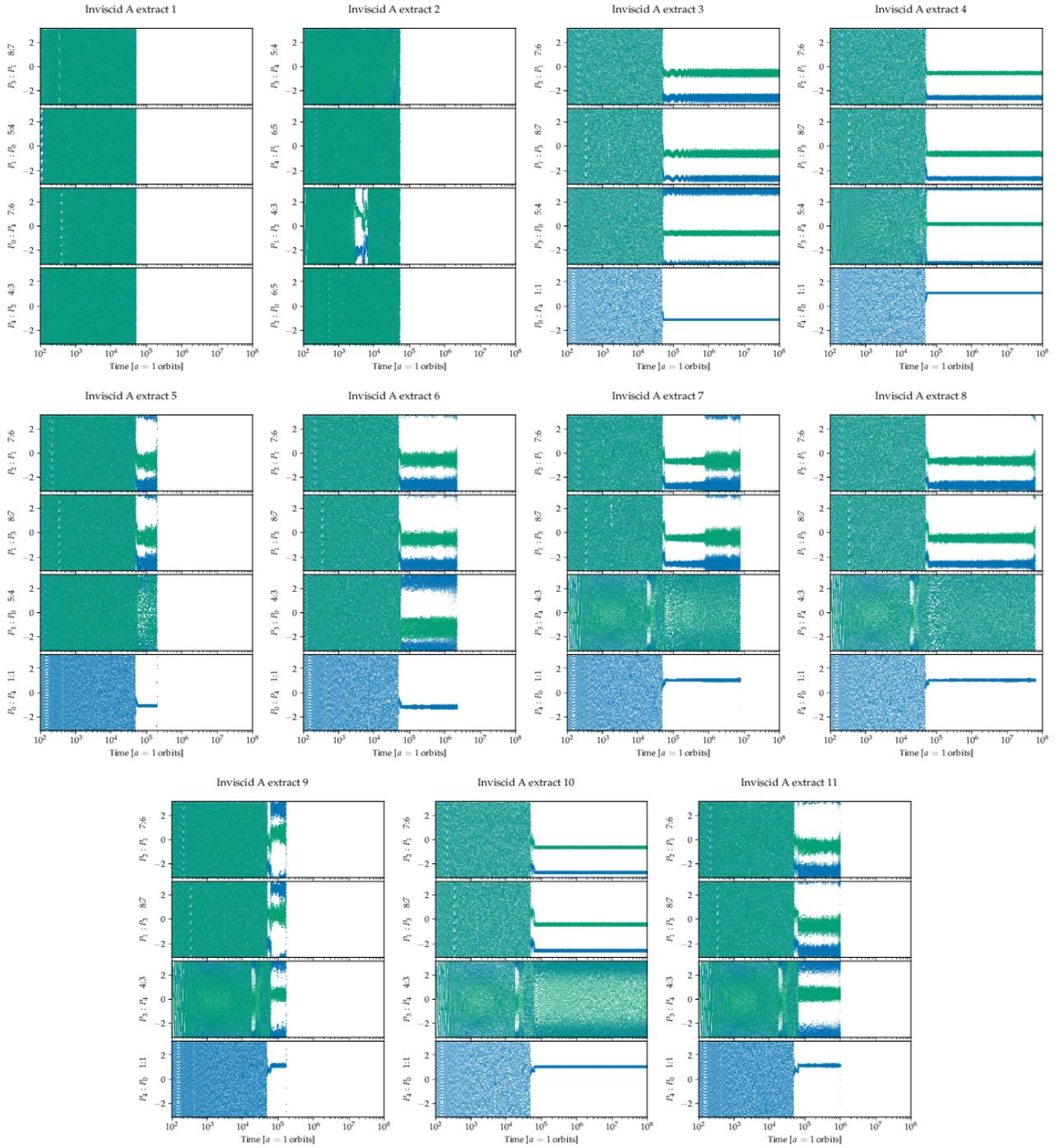

Figure 9: Nearest final first-order mean-motion resonant angles for run Inviscid A including disc dissipation and pure n-body phase.

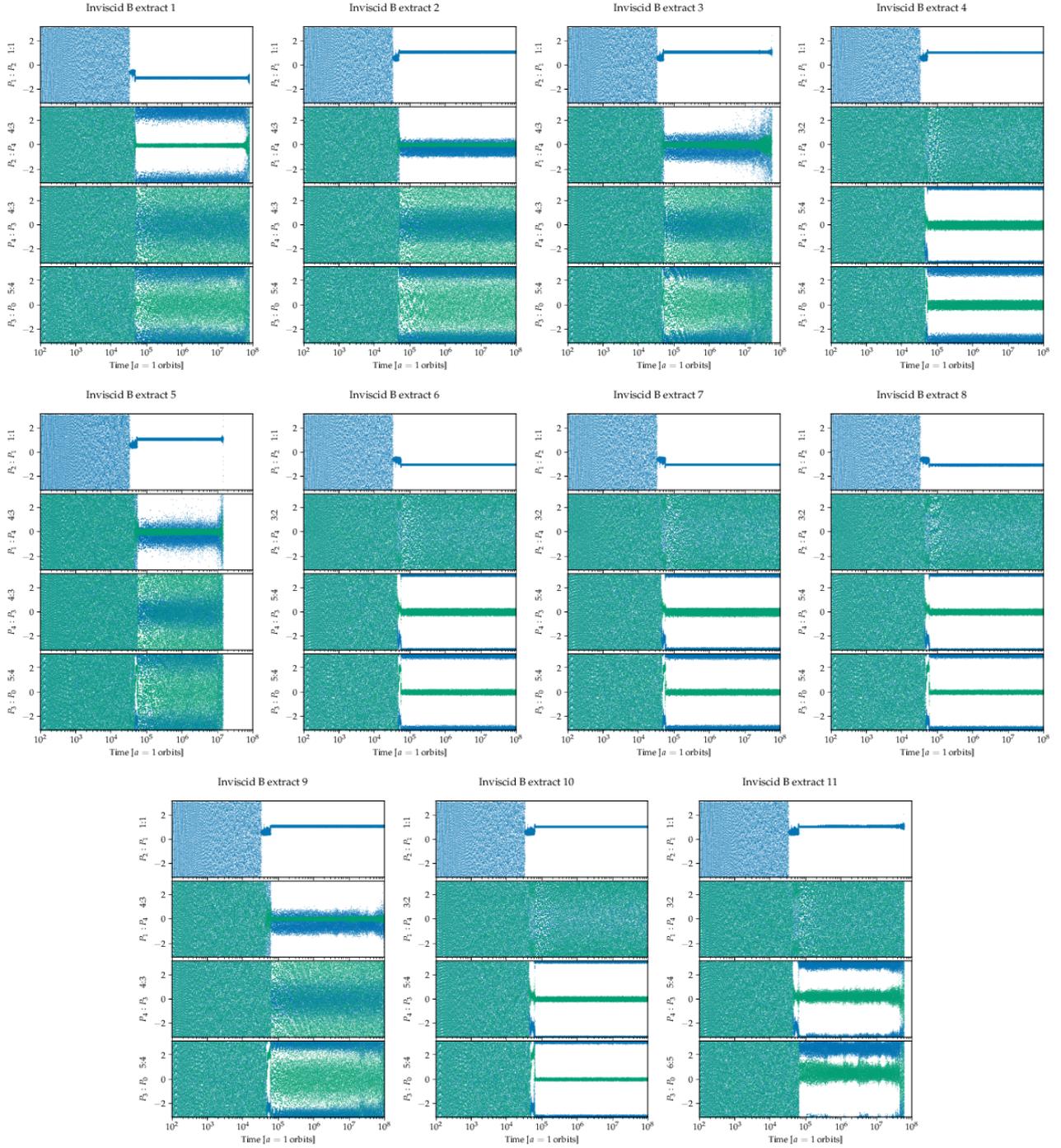

Figure 10: Nearest final first-order mean-motion resonant angles for run Inviscid B including disc dissipation and pure n-body phase.

12

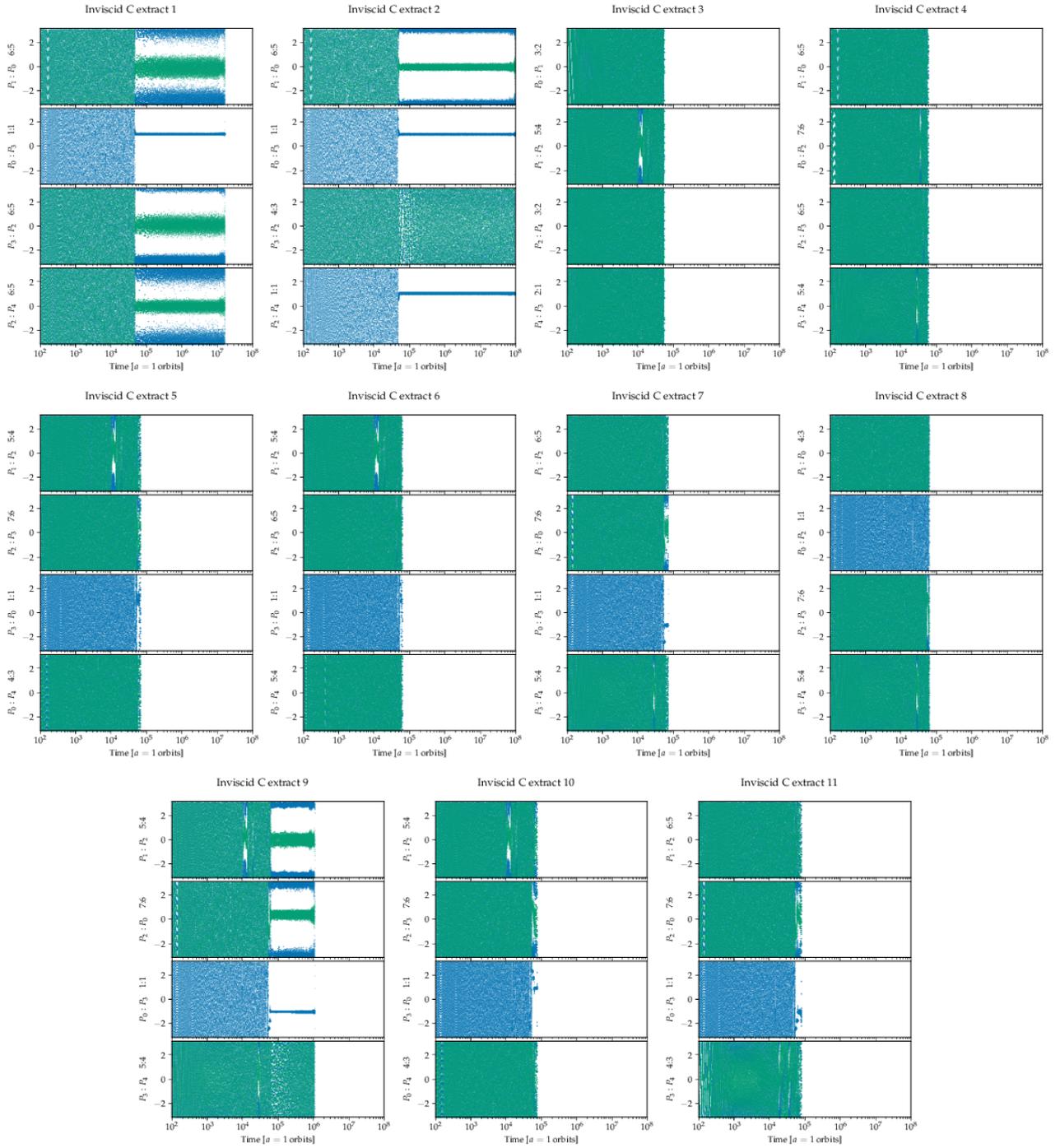

Figure 11: Nearest final first-order mean-motion resonant angles for run Inviscid C including disc dissipation and pure n-body phase.

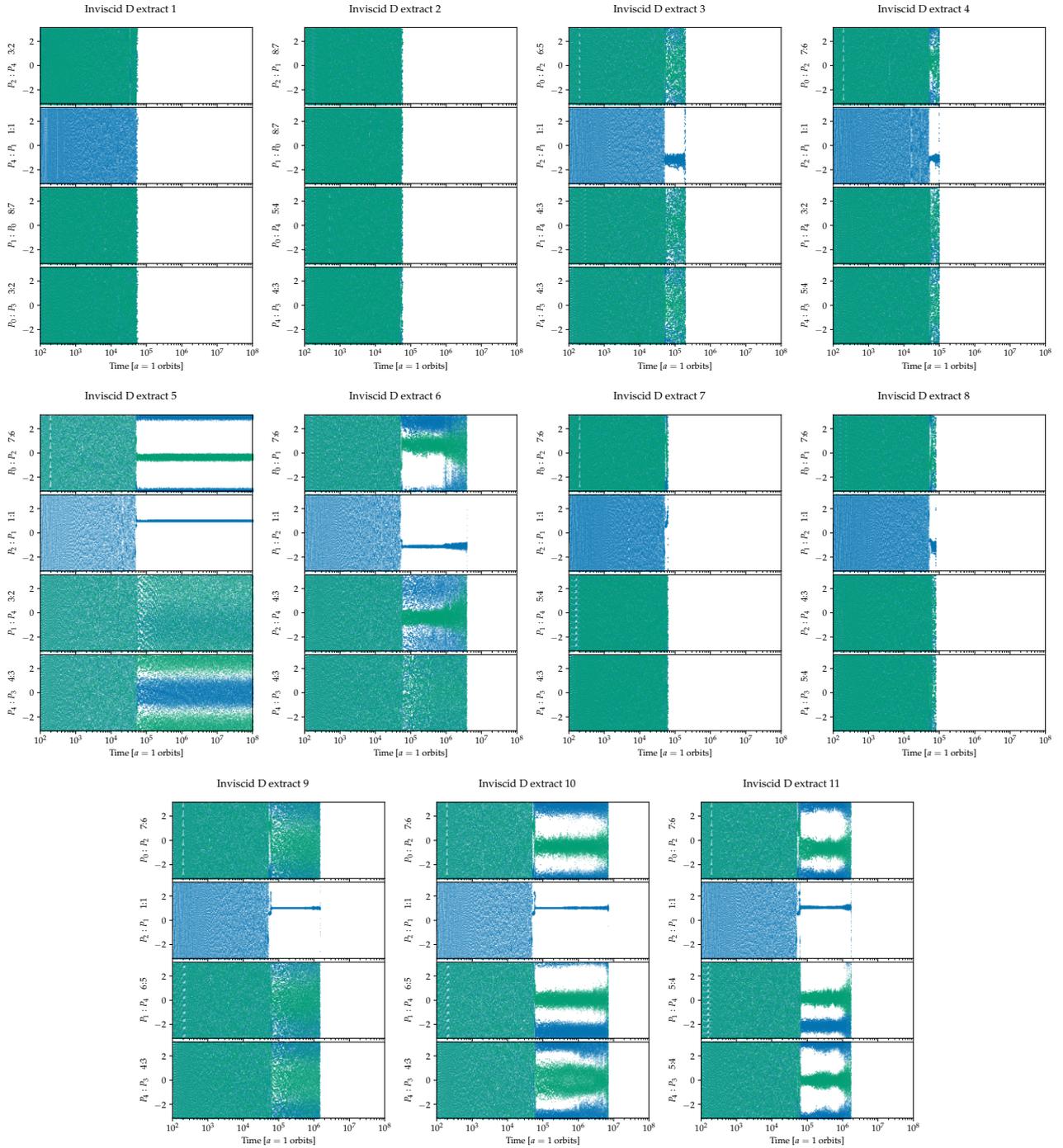

Figure 12: Nearest final first-order mean-motion resonant angles for run Inviscid D including disc dissipation and pure n-body phase.

14

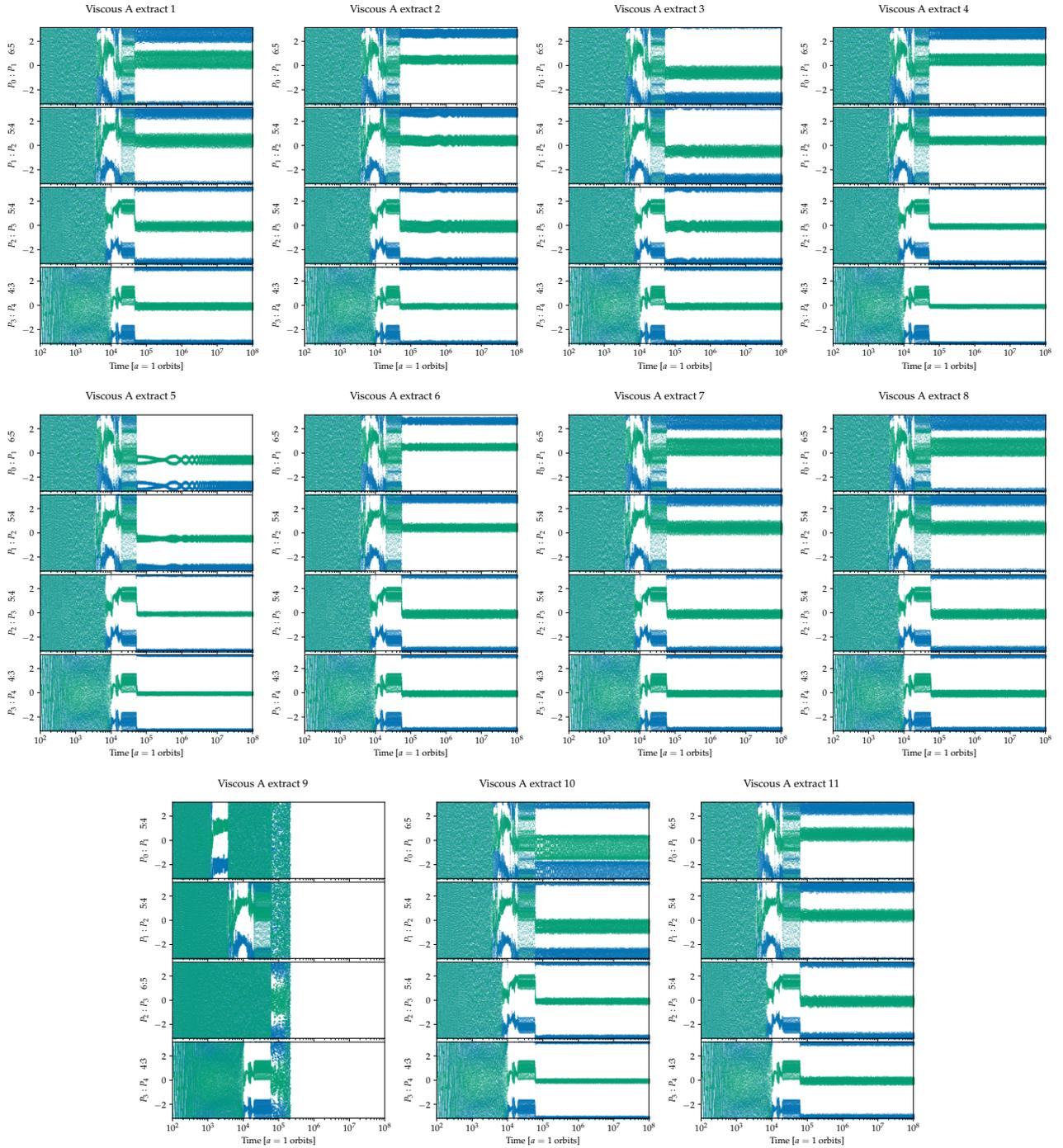

Figure 13: Nearest final first-order mean-motion resonant angles for run Viscous A including disc dissipation and pure n-body phase.

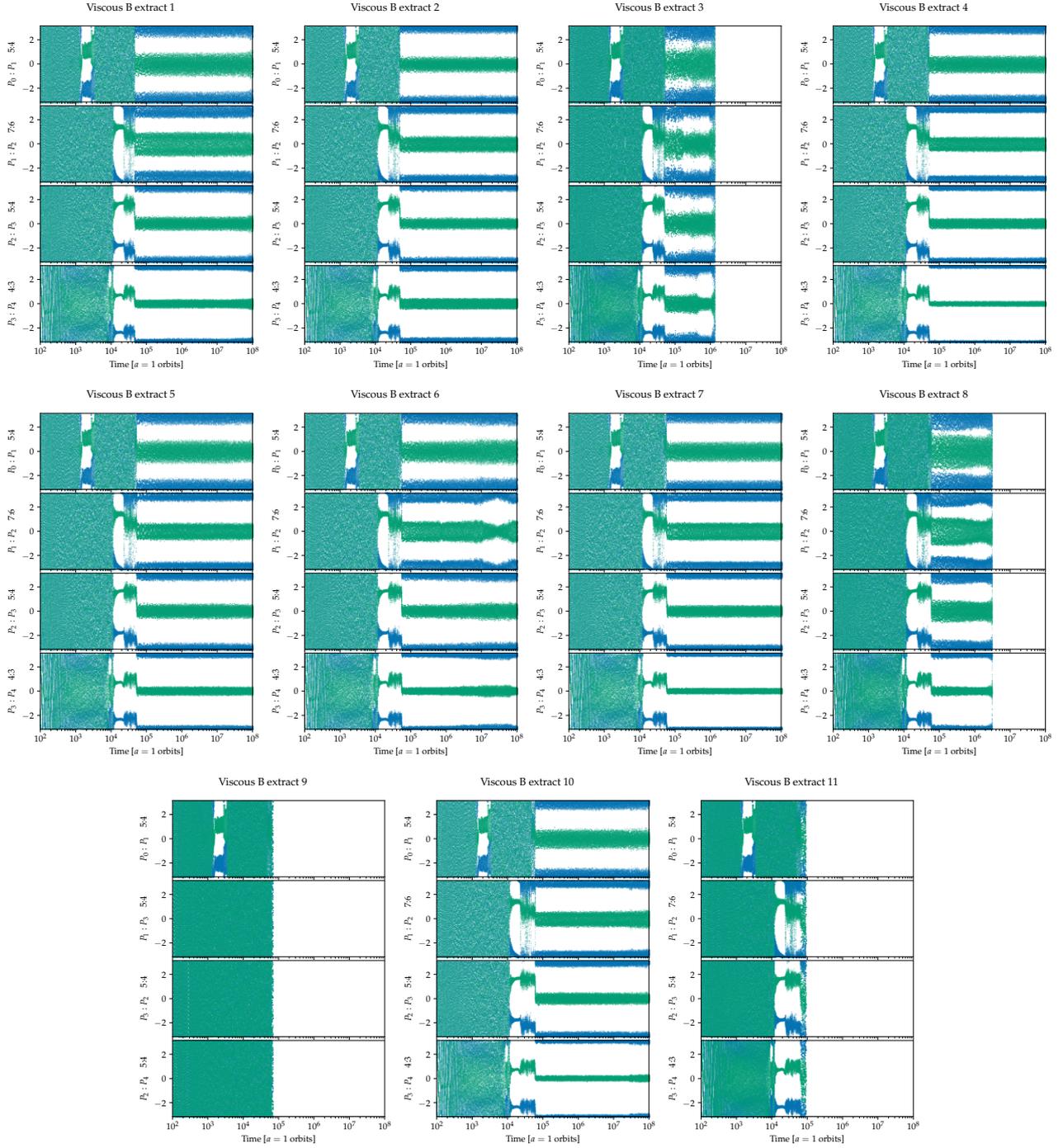

Figure 14: Nearest final first-order mean-motion resonant angles for run Viscous B including disc dissipation and pure n-body phase.

# D  Online Appendix: Sod shock tube verification test

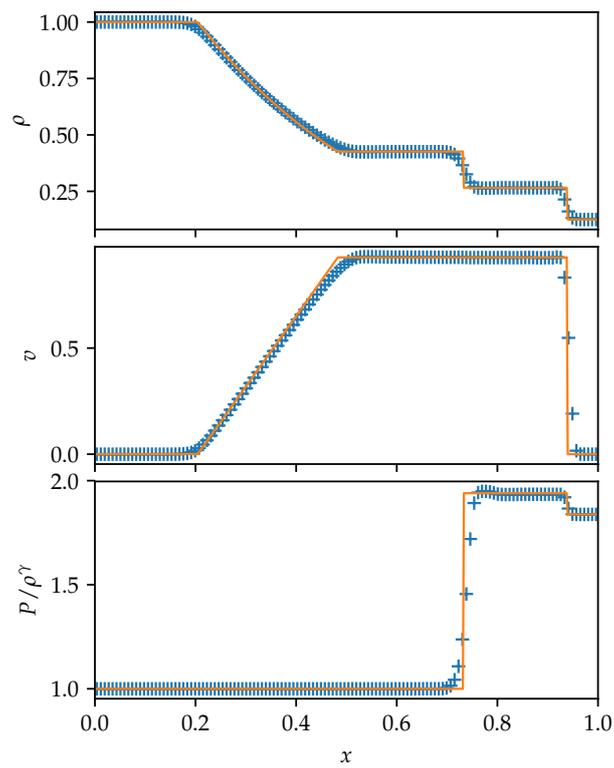

Figure 15: Sod shock tube test of the specific entropy formulation at $t = 0.25$ with 128 cells. Specific entropy is expressed in the conventional form $P/\rho^\gamma$. Analytical result shown in the solid orange line, solution values at each cell by + symbols.


\end{document}